# Cybersecurity Incident Response in Organisations: A Meta-level Framework for Scenario-based Training

## Full research paper

**Ashley O'Neil**
School of Computing and Information Systems
The University of Melbourne
Parkville, Victoria, Australia
Email: oneilla1@student.unimelb.edu.au

**Atif Ahmad**
School of Computing and Information Systems
The University of Melbourne
Parkville, Victoria, Australia
Email: atif@unimelb.edu.au

**Sean B. Maynard**
School of Computing and Information Systems
The University of Melbourne
Parkville, Victoria, Australia
Email: seanbm@unimelb.edu.au

## Abstract

Cybersecurity incident response teams mitigate the impact of adverse cyber-related events in organisations. Field studies of IR teams suggest that at present the process of IR is under-developed with a focus on the technological dimension with little consideration of practice capability. To address this gap, we develop a scenario-based training approach to assist organisations to overcome socio-technical barriers to incident response. The training approach is informed by a comprehensive list of socio-technical barriers compiled from a comprehensive review of the literature. Our primary contribution is a novel meta-level framework to generate scenarios specifically targeting socio-technical issues. To demonstrate the utility of the framework, a proof-of-concept scenario is presented.

# 1   Introduction

Organisations are under constant threat from highly-sophisticated attackers. The threat landscape is rapidly evolving and as technologies advance, attackers find new ways to penetrate organisations' digital defences in the hopes of obtaining information or to sabotage their operations (Ahmad et al. 2019). Much active research exists into how we can fortify these digital defences, but comparatively little exists into strengthening the process that steps in when things go wrong: incident response (IR).

Cybersecurity security incident response teams are at the forefront when digital defences fail, and must step in to mitigate the damage and restore services (Ahmad et al. 2020). But at present, the ability of incident response teams to respond to cyber-attacks is being hampered by a broad range of socio-technical issues (Nyre-Yu et al. 2019). Addressing these barriers can help to improve overall incident response in organisations. One way to do this is through training, in particular using scenarios in training, as they can improve team performance in cybersecurity through building skills and identifying potential weaknesses. This leads to the following research question:

*"How can scenarios be developed to improve cyber incident response in organisations?"*

The paper is structured as follows: the research methodology is first outlined for identifying the challenges being faced by incident response teams, before a literature review is conducted of the socio-technical perspective of IR. From here, existing research into scenario development is synthesised and critically analysed in the context of the research question. Next, a meta-level framework for scenario development in IR is proposed by drawing on the previous analysis, before providing a proof-of-concept scenario. Lastly, the implications of the research from the practice and theory perspectives is discussed.

# 2   Research Methodology

We systematically identified relevant contributions in journals and conference proceedings using (Webster and Watson 2002). We then used forward and backward chaining to identify further articles. The review focused on the study of real-world IR practice and the challenges faced by organisations. Google Scholar was used to carry out the search as it provided a systematic and efficient way to search online databases. In addition, it also allowed for the inclusion of grey literature and provided citation-specific information, such as how often an article had been cited and by whom.

The key search terms used were ("incident response") AND ("cyber security" OR "information security" OR "practice"). Each search was limited to the top 20 most relevant articles (as ranked by Google Scholar) as few articles were found to be of relevance after the top 10 and none after the top 20. Duplicates were then removed. The search resulted in 56 papers that were further analysed by the researcher by assessing the quality of each paper's source and the relevance of each article based on criteria. The results were limited to articles published only in peer-reviewed venues (40 papers); and those that met the criteria of both relating to cybersecurity and discussing IR capabilities in practice (13 papers). As the research focuses on current IR capabilities, only papers published in the past decade (2011-2021) were considered. This reduced the number of results to 10.

In the second part of the research, we investigated various approaches to scenario building for event-based training to determine the best methods for scenario building in IR. Subsequently, we adopted training aspects of two event-based approach to training (EBAT) methodologies (Nguyen et al. 2016; Oser et al. 1999) and the cybersecurity scenario design methodology (Guerber et al. 2010) to develop a framework for scenario development in IR. We then develop a proof-of-concept scenario.

# 3   Socio-technical issues facing organisational incident response

Although the importance of IR in organisations is widely acknowledged in literature, the discourse is largely technology-centric (Ahmad et al. 2021a). Comparatively less attention is given to socio-technical perspectives of how IR is managed in real-world practice. We abstracted the issues identified in the literature impacting organisational IR and grouped them according to the people, process, technology (PPT) framework of information systems.

Technology plays the role of both barrier and enabler to IR. As sophisticated attackers find new ways to penetrate the digital defences of organisations, the cybersecurity community hits back by developing new tools to detect and thwart them. But advancements in machine learning algorithms or artificial intelligence can create problems as well as solve them, and, as such, emphasis needs to be placed on how these new technologies integrate with the process and people in the organisation (Nyre-Yu et al. 2019). There is a need for better tools in IR as existing tools are highly useful but suffer from a high false positive rate and lack of usability (Tøndel et al. 2014). As a workaround, teams often create custom tools to achieve greater results. In addition, highly specific knowledge is required to interpret network traffic and use the tools, but was rarely documented. Issues also arose in integrating output from multiple monitoring tools to create a bigger picture and the level of complexity in an organisation's systems and services meant that incident responders were instead relying on notifications from people. These findings were backed by Ahmad et al. (2021a) who labelled this "poor visibility" as a key challenge and further identified a lack of optics in non-IT domains, such as HR. These issues affecting technology were often compounded by poor integration between technology, people and the processes of an organisation.

Organisational processes impact both technology and people, and vice versa. Nyre-Yu et al. (2019), note that the issues affecting IR are "more than just a set of usability issues in software, or of technology development and deployment" and instead span multiple levels of an organisational hierarchy, with each interconnected (p. 438). Examining three computer security IR teams Nyre-Yu et al. (2019) found capability was constrained by whether or not security was identified as a priority in the organisational mission. If not, it was seen as a competition to operational uptime and restricted growth and resources available to IR teams. This is backed by Ahmad et al. (2021a) who report on IR being further constrained by its positioning in IT support operations (and thus seen as a cost-center). Similarly, Tøndel et al. (2014) report information security (InfoSec) being "viewed merely as a technical issue" (p. 53). This view leads to a technology-centric focus of IR, concerned more with restoring IT services than any bigger picture at play (Ahmad et al. 2020; Ahmad et al. 2021a).

The structure of the IR team can further constrain capabilities. IR teams in Security Operations Centres (SOCs) are often segregated into "tiers" with members grouped together according to experience. This restricts the expertise available to each tier and hinders information sharing between analysts (Nyre-Yu et al. 2019) and leads to a culture that erodes collaboration among team members, favouring individual achievement over teamwork (Ahmad et al. 2021a). Furthermore, incidents are often escalated up the chain and sometimes out of the team altogether, limiting the authority of the IR team (Nyre-Yu et al. 2019). In the wider structure, relationships between team units can also impact IR capability. InfoSec management teams also appear alongside IR in large organisations. Ahmad et al. (2020) argue that these teams are often disconnected and have weak process-level integration, which leads to a fragmented approach to incidents. This is manifested in a lack of communication, collaboration, and knowledge sharing, which degrades IR capabilities and extends to other business units inside the organisation, which cumulatively have the capability to result in a "strategic-level disconnect". This disconnect was explained by Ahmad et al. (2020) with a hypothetical scenario whereby each unit in turn failed to provide the other with crucial business context (due to lack of communication, collaboration, and knowledge sharing), and ultimately ended up with a successful attack on the organisation.

From a people perspective, closely linked to the integration of teams and processes surrounding the handling of incidents, is the ability of teams to collaborate and communicate effectively. Knowledge or information sharing was a frequent issue impacting IR capabilities reported in the literature. The Ahmad et al. (2021b) case study is unique as it is the only one that labels its subject organisation as exhibiting "exemplar" practices. But a key reason why is because it is able to effectively share knowledge across its security team and wider to other IT and business units. This is not the case reflected in the research by Nyre-Yu et al. (2019), who cite information sharing as key to developing a shared awareness of incidents across IR teams, IT, and the wider organisation. Tøndel et al. (2014) and Ahmad et al. (2020) also cite issues with information sharing, with personnel unsure of what they should report and who they should report it to; and a lack of formal policies on what should be shared and which communication channels to use. A poor organisational culture of mistrust between teams and fear over reporting incidents for who would be held responsible, were also acknowledged as fuelling poor communication and collaboration (Tøndel et al. 2014). Outsourcing of services created further barriers with suppliers often excluded in many phases of IR, or unwilling to take responsibility (Tøndel et al. 2014).

Training and formal policies for effective communication and knowledge sharing are cited as ways to overcome the barriers outlined above, but they were rare in practice (Tøndel et al. 2014). Furthermore, individual awareness of InfoSec was an issue, as training tended to focus on technical staff, not all employees — this becomes an issue in IR as people (and their resulting notifications) were relied on heavily in the literature in detection of incidents (Tøndel et al. 2014). In addition, more training was required for skills beyond that of the technical nature, as they were crucial in cybersecurity (Van der Kleij et al. 2017).

Steinke et al. (2015) focused on these "soft skills" and reflect on the environment that incident responders operate in – dynamic and uncertain – and outline numerous non-technical skills that are integral to a team's efficiency, such as adaptation, communication, problem solving, and trust. Van der Kleij et al. (2017) echoed this, listing "social skills" as a need for incident responders, but gaps in these skill sets existed in practice. Both studies by (Tøndel et al. 2014) and Nyre-Yu et al. (2019) raise the issue of expertise of staff, with barriers such as budget and organisational priorities impacting a team's ability to upskill. Consequently, staff find workarounds, relying on manual detection or the improper use of tools to get the job done (Tøndel et al. 2014).

The identified issues outlined above are seen in the available literature as the major challenges to IR capabilities. It is however important to note these issues are constrained to what is presented in the research on IR, and often it is not from exemplar organisations. Furthermore, as highlighted by Tøndel et al. (2014), the data collection methods vary widely and the absence of an identified practice does not necessarily mean it is not performed, only undocumented.

## 4  Framework Development: Scenario building

Scenario-based training (SBT) is a term that is at times well-defined but also more loosely applied to any form of training using a scenario as the curriculum (Oser et al. 1999). SBT uses the scenario itself as a training tool to develop and test participants' responses to the situation unfolding (Moats et al. 2008). Methodologically, SBT includes a scenario design phase, the delivery of the training session and debriefing afterwards (Moats et al. 2008). A critical aspect is to start by identifying learning objectives for the training event that are observable and measurable for later stages (Cannon-Bowers 2008). One approach is EBAT (event-based approach to training), which takes the process further and systematically introduces exercise events that map to these learning objectives, as well as participant feedback (Fowlkes et al. 1998; Nguyen et al. 2016). These linkages allow training participants to demonstrate any proficiencies (or deficiencies) in the learning objectives and allow facilitators to measure their performance and provide future learning opportunities through feedback (Fowlkes et al. 1998).

SBT is relevant to the research question in several ways. First, the exercise is largely driven by the identified learning objectives set at the beginning, which are then mapped to trigger events in the scenario. This provides an opportunity to "set the scene" with the issues identified in section 3. Second, the methodology then allows for organisational learning in the latter stages when participants' performance is measured, and feedback is offered to improve overall performance. This speaks to the research goal of improving organisational IR. Third, both SBT and EBAT are widely used to develop and test team performance (Fowlkes et al. 1998; Oser et al. 1999). Teamwork was a critical factor in how well an IR team operated in much of the literature from section 3. And, lastly, SBT is primarily focused on *training*, that is, how the scenario can help organisations improve their reaction to a situation by training and testing participants' responses to it (Cannon-Bowers 2008; Moats et al. 2008). This is a critical component of this paper's aim - improving organisational IR through scenario development.

SBT is already evident in cybersecurity and IR, where scenarios feature heavily in best practice training guides (Guerber et al. 2010; Kick 2014). However, they rarely offer much in terms of how to develop a scenario, instead more concerned with how the training exercise is conducted. (Guerber et al. 2010) is the exception and the method outlined is the most rigorous identified for developing cybersecurity scenarios. The three phases of scenario development provide a highly-detailed methodology and cumulatively is a process unlike any other identified in this research.

We therefore draw the conclusion that SBT, and more specifically, EBAT, are highly applicable to this research. Furthermore, the method proposed by (Guerber et al. 2010) is useful in developing scenarios, but in isolation, none of these approaches fulfill the research goal of developing scenarios to address the socio-technical barriers being faced by organisational IR.

### 4.1  Towards a framework for scenario development in incident response

We propose an adapted six-step framework (Table 1) to allow organisations to develop highly-tailored scenarios of cyber-attack, but to also improve IR capabilities by utilising these scenarios in an encompassing training program. The goal of this research is to develop scenarios that can improve IR capabilities by focusing on the socio-technical issues being faced by organisations. To this end, training aspects of two EBAT methodologies (Nguyen et al. 2016; Oser et al. 1999) have been combined with the unique cybersecurity scenario design methodology of Guerber et al. (2010). In doing so, the resulting scenarios (and encompassing training programs) draw on the SBT foundation of using scenarios to help organisations improve their reaction to a situation by training and testing participants' responses to it.

# 5   Findings: Applying the framework to develop a scenario

Drawing on the deficiencies in current practice outlined in section 3, we create a scenario derived from the framework in section 4 to help organisations improve their IR capabilities. The scenario will be the basis for a training exercise as per the framework, but for this paper we only focus on the scenario creation, not the training exercise. As such, the steps directly related to the exercise will be omitted (steps five and six). Furthermore, the scenario will be created to be as generic as practical to allow for increased utility in organisations, but at the same time will require some constraints, namely that it is designed for organisations with large heterogeneous technology estates, spanning multiple jurisdictions. Lastly, the scenario will be developed as a tabletop exercise (TTX), one of two common scenario-based training exercises in cybersecurity. A TTX was chosen as they are typically much shorter in duration than their functional exercise counterparts and require less planning and technical resources (Kick 2014).

| Scenario-development Steps | Scenario-building Activities |
|---|---|
| **Step 1:** Develop learning objectives | Identify the goals of the training program |
| **Step 2:** Craft trigger events | Provide opportunities for participants to demonstrate proficiencies / deficiencies for all learning objectives |
| **Step 3:** Develop scenario storyline | |
| **3.1:** Determine key scenario elements | Determine scenario intent, threat, target, operational effect, business impact |
| **3.2:** Develop backstory | Develop detail of threat actors, necessary intelligence and background information |
| **3.3:** Finalise storyline | Revisit each scenario element and add in any extra storyline details |
| **Step 4:** Develop event threads | |
| **4.1:** Craft event synopsis | Craft an event synopsis by outlining the chronological event thread that will stimulate storyline |
| **4.2:** Craft events | Fill in event details from previous step |
| **4.3:** Event thread walkthrough | Walkthrough to flesh out final scenario details |
| **Step 5:** Identify targeted responses to events and performance measures | Identify target responses to events or measures to observe performance for evaluation and feedback |
| **Step 6:** Operationalise learnings | Put learnings from exercise into practice to improve IR |

*Table 1. Scenario Development Framework*

## 5.1   Step 1: Develop learning objectives

As this example focuses on the creation of a scenario, and not the encompassing training exercise, the issues identified from the literature review (summarised in Table 2) are used in place of learning objectives. The identification of these issues is the cornerstone of this research and by explicitly linking them to trigger events, the scenario created will exploit any deficiencies in these areas, thus allowing an organisation to assess their IR capabilities. It is worth noting that these issues would have been turned into learning objectives if the full training framework was being followed.

## 5.2   Step 2: Craft trigger events

In this step, we map learning objectives (or IR issues) to trigger events to be used in the scenario. Trigger events were identified for each issue, by drawing on the literature and our knowledge of real-world cyber incidents. This linkage between the trigger events and issues will allow the resulting scenario to expose any weaknesses that may exist in these areas. Table 2 shows the outcome of this process, with the learning objectives or issues identified in section 3 listed in the left-hand column, along with the trigger events created in this step listed in the right-hand column. The list of trigger events is not exhaustive but is a useful starting point in crafting training scenarios that address deficiencies in "best practice" IR.

## 5.3   Step 3: Develop scenario storyline

The next step of the framework has been adapted from the cybersecurity training exercise methodology proposed by Guerber et al. (2010) and involves three sub-phases: determine key scenario elements; develop backstory; and finalise storyline.

### 5.3.1 Step 3.1: Determine key scenario elements

In this sub-phase, the key scenario elements of scenario intent, threat, target, and operational effect are determined. A fifth key element, "business impact", is introduced which is purposely excluded by Guerber et al. (2010) but is relevant here as the context for the scenario is cybersecurity in *organisations*. The *scenario intent* is described by Guerber et al. (2010) as the overall objective of the scenario. For this research, the scenario intent is twofold: to develop a scenario that exploits the issues impacting IR capabilities as identified in section 3; and to use such a scenario in a training exercise to help organisations to improve their IR capabilities (although this paper does not cover this).

| IR Issue | Trigger Event |
|---|---|
| Technology complexity | A. attack affects a high number of diverse devices |
| Poor field of vision | B. attackers employ high volume of known attacks<br>C. contradictory notifications of attack among IR tools |
| Lack of appropriate tools | D. multipronged attack whereby later attacks offer cover for earlier ones, deleting logs / evidence<br>E. attacks part of infrastructure where IR team does not have full optics because of inappropriate tools<br>F. attack has a physical aspect |
| Organisational positioning in IT operations | G. attackers target non-IT service or asset, such as the business side or a business asset, a physical domain, or HR |
| Segregated nature of IR team | H. incident responder working in isolation and does not seek team collaboration to thwart attack<br>I. attackers employ high volume of known attacks |
| Weak process-level integration between teams | J. attackers target obscure business asset |
| Poor fit between process and incident | K. attackers misdirect IR team to cause a misdiagnosis of incident classification |
| Lack of documentation when reporting, handling, and following up incidents | L. un- or inadequately documented information becomes relevant to IR team |
| Poor intra- and inter-team collaboration and comms | *tested in other trigger events* |
| Insufficient training and development of security awareness | M. attacker targets end user, who encounters problem they don't perceive as a threat and logs incident through help desk<br>N. attackers target employee's personal device<br>O. attack duration exceeds 24 hours requiring incident responders to cope with prolonged state of "emergency"<br>P. attackers target shadow IT |
| Lack of focus in developing soft skills | Q. attack puts strain on relationship between IR and business units by targeting business only |
| Lack of technical expertise in IR teams | R. IR team required to conduct forensic evidence collection |

*Table 2. Mapping between IR issue and trigger events*

The *threat* element encompasses both the actor and their method of attack. For this scenario, a highly-sophisticated and organised attacker, who belongs to an advanced persistent threat (APT) group, was chosen over the selection of a low-skilled attacker as that would have limited the applicability of the scenario to address the range of issues identified. As they belong to an APT group they also represent the most formidable threat (Ahmad et al. 2019) and would use both known exploits and zero-day attacks in a prolonged endeavour. As this scenario is a TTX and is designed to be as generic as practical to allow for increased utility in organisations, all key scenario elements are discussed only in high-level detail. Some constraints are needed to derive a meaningful scenario, and as such, the scenario has been created from the perspective of a large organisation with a core research and development (R&D) function. An R&D organisation is an example of an organisation where it not only has IT services that could be the target of a cyber-attack, but also intellectual property. In this way, the resulting scenario will exercise as many identified deficiencies as possible. It also means we can update the threat actor to be a nation state threat actor or a competitor to the organisation — or more than likely, both.

Similarly, the *target* from a high-level perspective is the intellectual property of the organisation, but numerous intermediary assets and services along the way will be affected in the prolonged attack, such as shadow IT, servers, and other business assets. It is difficult to discuss targets in any finer detail, as the organisation being targeted is purely hypothetical.

*Operational effect* and *business impact* are clearly separated by Guerber et al. (2010) with the former relating to the effect on the target and organisation, with respect to the CIA triad of confidentiality, integrity, and availability. For this scenario, as the organisation's intellectual property is the target, the effect then becomes a total loss of confidentiality, and a disruption in availability of IT services as a secondary vector in the attack. For business impact, Guerber et al. (2010) define it as the effect on the organisation's ability to carry out its mission. In this instance, the scenario has been created from the point of view of an organisation whose primary function is R&D, and as such, the theft of its intellectual property would be catastrophic, resulting in the destruction of its competitive advantage at the least.

### 5.3.2　Step 3.2: Develop backstory

Guerber et al. (2010, p. 34) describe this sub-phase as the general context or "state of the world" for the scenario and it encompasses threat actor details and necessary background information on the organisation and nation. For the context of this scenario, only high-level detail will again be provided.

The *threat actor detail* comprises both motives and level of expertise. As determined in the previous section, the threat actor belongs to an APT group and is operating with a very high-level of expertise. In addition, the attack is nation-state sponsored and will be carried out with the motive of stealing the organisation's intellectual property to get financial or commercial advantage. The *necessary background information* can be viewed from both inside the organisation and outside. Internally, the scenario is presented during a time of considerable growth for the organisation, where its R&D operation has expanded significantly, and remains busy. This has resulted in the organisation employing new staff across the board and hiring contractors to upgrade systems in a timely manner to ensure minimal downtime. Externally, there has been plenty of interest in the organisation's newest developments in R&D, from both competitors and potential buyers. At the same time, the threat landscape is rapidly evolving, with highly-skilled attackers increasingly going undetected.

### 5.3.3　Step 3.3: Finalise storyline

This sub-phase involves revisiting each scenario element to add in any extra storyline details. Examples of extra storyline details include locations of specific systems, information or processes being targeted; the methods of discovery employed by the attacker to access a target; and an attacker's source location (see Guerber et al. (2010, p. 34) for more examples). Once again, as the scenario being developed here is based on a hypothetical organisation, and created to be as generic as practical, it is difficult to provide such granular detail. As such, no extra information will be added to the scenario elements.

## 5.4　Step 4: Develop event threads

The next step of the framework has also been adapted from the methodology proposed by Guerber et al. (2010) and again also encompasses three sub-phases, but was designed for functional exercises only, and not discussion-based ones like TTXs. As such, we will only adopt certain aspects of each sub-phase. The three sub-phases are: craft event synopsis; craft events; and event thread walkthrough. Cumulatively, the sub-phases are designed to develop event threads, which are chronological sets of events that relate to a specific focus area of the training program.

### 5.4.1　Step 4.1: Craft event synopsis

The first sub-phase, as outlined by Guerber et al. (2010), is to craft an event synopsis, or multiple synopses, in the case where an exercise has multiple cyber incidents. Each event synopsis provides a chronological list of all the events that make up the cyber incident (see Guerber et al. (2010, p. 36)). As much of this phase involves the technical detail required of functional exercises, it is inapplicable here. But at a high-level, it is still possible to use the list of trigger events identified in Table 2 to craft event synopses for the scenario being created. (Note, the scenario will have multiple related cyber incidents, and thus multiple synopses, in an effort to include as many of the trigger events in Table 2 as practical. It would be infeasible to incorporate them all in one cyber incident, and as such many would be missed, in turn reducing the number of issues that are exercised as each trigger event is mapped to at least one of the identified deficiencies in IR practice.)

To craft these event synopses, selected trigger events were clustered together to form a plausible cyber incident. These trigger events were then arranged in the order they would expect to be seen in the determined cyber incident. The outcome is shown in Table 3, where each event synopsis represents one cyber incident that will be observed in the scenario.

| Trigger Event | Story Line |
|---|---|
| **Event synopsis 0** | |
| G. attackers target non-IT service or asset, such as the business side or a business asset, a physical domain, or HR<br>F. attack has a physical aspect | • a new hire at the organisation is actually working for the attackers<br>• attackers gain physical access to systems being upgraded by posing as contractors |
| **Event synopsis 1** | |
| P. attackers target shadow IT<br>N. attackers target employee's personal device<br>M. attacker targets end user, who then encounters problem they don't perceive as a threat and logs incident through help desk<br>L. un- or inadequately documented information becomes relevant to IR team<br>R. IR team required to conduct forensic evidence collection | • attackers compromise R&D worker's personal device in the hunt for intellectual property<br>• R&D worker does not realise as they lack InfoSec awareness knowledge and instead log resulting technical issues through help desk<br>• IR team eventually detects the incident but must deal with lack of information and processes around personal devices |
| **Event synopsis 2** | |
| G. attackers target non-IT service or asset, such as the business side or a business asset, a physical domain, or HR<br>J. attackers target obscure business asset<br>K. attackers misdirect IR team to cause a misdiagnosis of incident classification<br>Q. attack puts strain on relationship between IR and business units by targeting business only<br>R. IR team required to conduct forensic evidence collection | • attackers take R&D server offline (along with others) to hide the theft of intellectual property<br>• IR team must decide between focusing on restoration or exploring the root cause of the incident |
| **Event synopsis 3** | |
| B. attackers employ high volume of known attacks<br>I. attackers employ high volume of known attacks<br>C. attack results in contradictory notifications among IR tools<br>O. attack duration exceeds 24 hours requiring incident responders to cope with prolonged state of "emergency" | • attackers launch high-volume attack designed to confuse and fatigue incident responders, and provide cover via misdirection/interference<br>• IR team cannot cope with the sustained increase in volume and it impacts team culture |
| **Event synopsis 4** | |
| A. attack affects a high number of diverse devices<br>C. attack results in contradictory notifications among IR tools<br>D. attack is multipronged, whereby later attacks offer cover for earlier ones, deleting logs and forensic evidence | • attackers launch large-scale attack, e.g. distributed denial of service, to wipe any evidence left behind<br>• notifications confuse incident responders as to the motives of the attack, with tools giving them conflicting messages<br>• IR team must decide between focusing on restoration or exploring the root cause of the incident |

*Table 3. Event synopsis*

It is possible for two or more trigger events to occur at the same time in each cyber incident, but for the purposes of the table, they are still listed one after each other. Of the 18 trigger events identified in Table 2, all but two have been included in the cyber incidents to keep the resulting scenario both plausible and to an acceptable length. The two that were excluded, however, map to issues that are already represented in the included 16 trigger events, so no issue has been left out of the resulting scenario.

#### 5.4.2    Step 4.2: Craft events

The second sub-phase exists to fill in the event details from the previous step. Keeping with a high-level discussion, for each event synopsis outlined in Table 3, a storyline has been crafted to utilise the respective event triggers.

### 5.4.3    Step 4.3: Event thread walkthrough

The third and last sub-phase is a walkthrough of the scenario to flesh out the final details. It is a team exercise and Guerber et al. (2010) identify it as the most time-consuming phase of scenario development. Expected responses from players are also included in this sub-phase, however, for the proposed framework, it is included in the following step – step 5, identify targeted responses to events and performance measures – in keeping with the EBAT methodologies.

In the context of the scenario being developed here, the storylines have been "fleshed out" with finer details to create a cumulative scenario of five incidents or event threads. The scenario has been presented as a TTX, consistent with best practice guidelines (Guerber et al. 2010; Kick 2014) and examples of TTXs found online (e.g. WSOC (2014)). The resulting TTX is included in Appendix A.

## 6    Discussion

Returning to the research question of "how can scenarios be developed to improve cyber incident response in organisations?", the first part of the project identified a comprehensive list of socio-technical issues facing organisational IR through the systematic literature review. The second part of the research identified a lack of an appropriate methodology to develop scenarios to address these issues, and as such, proposed a new framework that is rooted in the previous work of EBAT and best practice cybersecurity guidelines. A scenario was then created from this framework, which systematically mapped the socio-technical issues identified in the first part of the research to events in the resulting scenario. Owing to the previous work in the utility of EBAT and SBT overall, it is argued that such a scenario would not only enable organisations to assess their IR capabilities but also improve them by following the extra steps to implement the encompassing training program.

This research is significant to both the practice and theory of IR. The meta-level framework developed, the proof-of-concept scenario, the comprehensive list of socio-technical issues, and the event triggers derived from them, all take strides to address a real-world problem. That is, organisations currently lack the capability to carry out effective IR because the process is technology-focused and under-developed. By focusing on the socio-technical barriers we argue organisations can improve their IR capability.

This research is significant to practice because it provides organisations with a ready-made scenario to improve IR capability and teaches them how to create their own and use it. The framework developed and the two "ingredients" created in this project – the list of IR deficiencies and the event triggers derived from them – combine into a training tool that, to the best of our knowledge, is the first of its kind in cybersecurity. It targets the socio-technical barriers we identified as currently impacting IR teams, allowing organisations to assess whether they are facing the same issues. But the framework doesn't solely rely on these barriers — it has been created in a way as to be flexible enough to allow organisations to choose specific deficiencies to focus on, or they can create their own clearly crafted learning objectives entirely. The encompassing training program will lead to better aligned outcomes for IR and all six areas of InfoSec management: IR; policy; risk management; education, training, and awareness; technical management; and intra-organisation liaison management (Alshaikh et al. 2014). From a theory perspective, the research addresses a gap in the literature with no comparable methodology identified for organisations to develop scenarios (and resulting training programs) that target socio-technical issues. The training framework draws on established methodologies to systematically link the socio-technical issues with events in the resulting scenario. The comprehensive range of deficiencies we identified consolidates the available literature on IR practice, which is at present dispersed and disconnected. New research in IR is needed — especially regarding the study of IR in practice. Furthermore, IR literature that focuses on socio-technical issues is lacking, with the focus currently mainly on technology aspects of IR.

The foundation of this research is the literature contribution (synthesised in the first phase). The results are influenced by not only what was *available*, but also the *scope* of such literature and the *context* of the organisations studied (e.g. typically not exemplar organisations). More research is needed into the real-world practice of IR to remedy this. Furthermore, data collection methods vary widely in the available literature, and the absence of an identified practice does not necessarily mean it is not performed, only undocumented (Tøndel et al. 2014).

Additionally, the project was subject to several limitations. Firstly, the scenario was developed from the viewpoint of a hypothetical rather than an actual organisation. Secondly, the encompassing training program was omitted from the proof of concept, instead with the focus solely on the scenario creation. From here, the framework needs to be tested by organisations, who would be able to implement it in full, creating not only a scenario driven by socio-technical issues, but an encompassing training program to operationalise learnings from the scenario. The results would enable researchers to measure the effectiveness of the framework to see if it in fact could help improve organisational IR.

In addition, the research presents several other opportunities for future work. Firstly, more research is needed into the practice of IR. More case studies examining the real-world practice of IR teams would go a long way to furthering this research into the socio-technical challenges faced in organisational IR. It could do this by both confirming or denying the presence of the issues identified in this literature review, and also identifying new issues. Furthermore, greater focus is also needed into the socio-technical perspectives of IR. Secondly, the framework could be extended for use in functional cybersecurity exercises as well as discussion-based ones like TTXs. Lastly, the suggested list of event triggers could be enhanced through greater research into past cyber-attacks, bringing more depth to the scenarios generated as a result.

## 7  Conclusion

How well an organisation detects, contains, and recovers from a cyber-attack ultimately relies on the capability of its IR team. IR teams face a multitude of socio-technical barriers that are often the underlying cause of underperformance. We developed a training tool that will enable organisations to identify their specific barriers as a baseline for improvement.

## 8  References

Ahmad, A., Desouza, K. C., Maynard, S. B., Naseer, H., and Baskerville, R. L. 2020. "How Integration of Cyber Security Management and Incident Response Enables Organizational Learning," *Journal of the Association for Information Science and Technology* (71:8), pp. 939-953.

Ahmad, A., Maynard, S. B., Desouza, K. C., Kotsias, J., Whitty, M. T., and Baskerville, R. L. 2021a. "How Can Organizations Develop Situation Awareness for Incident Response: A Case Study of Management Practice," *Computers & Security* (101), p. 102122.

Ahmad, A., Maynard, S. B., Motahhir, S., and Anderson, A. 2021b. "Case-Based Learning in the Management Practice of Information Security: An Innovative Pedagogical Instrument," *Personal and Ubiquitous Computing*).

Ahmad, A., Webb, J., Desouza, K. C., and Boorman, J. 2019. "Strategically-Motivated Advanced Persistent Threat: Definition, Process, Tactics and a Disinformation Model of Counterattack," *Computers & Security* (86), pp. 402-418.

Alshaikh, M., Ahmad, A., Maynard, S. B., and Chang, S. 2014. "Towards a Taxonomy of Information Security Management Practices in Organisations," *25th Australasian Conference on Information Systems*, Auckland, New Zealand, p. 10.

Cannon-Bowers, J. A. 2008. "Recent Advances in Scenario-Based Training for Medical Education," *Current Opinion in Anesthesiology* (21:6), pp. 784-789.

Fowlkes, J., Dwyer, D. J., Oser, R. L., and Salas, E. 1998. "Event-Based Approach to Training (Ebat)," *The international journal of aviation psychology* (8:3), pp. 209-221.

Guerber, A., Fogle, C., Roberts, C., Evans, C., MacDougald, B., and Butkovic, M. 2010. "Methods for Enhanced Cyber Exercises." Carnegie Mellon University.

Kick, J. 2014. "Cyber Exercise Playbook," MITRE Corp, Bedford MA, USA.

Moats, J. B., Chermack, T. J., and Dooley, L. M. 2008. "Using Scenarios to Develop Crisis Managers: Applications of Scenario Planning and Scenario-Based Training," *Advances in Developing Human Resources* (10:3), pp. 397-424.

Nguyen, N., Watson, W. D., and Dominguez, E. 2016. "An Event-Based Approach to Design a Teamwork Training Scenario and Assessment Tool in Surgery," *Journal of surgical education* (73:2), pp. 197-207.

Nyre-Yu, M., Gutzwiller, R. S., and Caldwell, B. S. 2019. "Observing Cyber Security Incident Response: Qualitative Themes from Field Research," *Proceedings of the Human Factors and Ergonomics Society Annual Meeting*: SAGE Publications Sage CA: Los Angeles, CA, pp. 437-441.

Oser, R. L., Gualtieri, J. W., Cannon-Bowers, J. A., and Salas, E. 1999. "Training Team Problem Solving Skills: An Event-Based Approach," *Computers in human behavior* (15:3-4), pp. 441-462.

Steinke, J., Bolunmez, B., Fletcher, L., Wang, V., Tomassetti, A. J., Repchick, K. M., Zaccaro, S. J., Dalal, R. S., and Tetrick, L. E. 2015. "Improving Cybersecurity Incident Response Team Effectiveness Using Teams-Based Research," *IEEE Security & Privacy* (13:4), pp. 20-29.

Tøndel, I. A., Line, M. B., and Jaatun, M. G. 2014. "Information Security Incident Management: Current Practice as Reported in the Literature," *Computers & Security* (45), pp. 42-57.

Van der Kleij, R., Kleinhuis, G., and Young, H. 2017. "Computer Security Incident Response Team Effectiveness: A Needs Assessment," *Frontiers in psychology* (8), p. 2179.

Webster, J., and Watson, R. T. 2002. "Analyzing the Past to Prepare for the Future: Writing a Literature Review," *MIS Quarterly* (26:2), pp. xiii-xxiii.

WSOC. 2014. "Tabletop Exercises," Washington State Office of Cybersecurity, Washington Technology Solutions. .

# 9    Appendix A: Proof-of-Concept Scenario

Preamble: Your organisation has just gone through a period of considerable growth within its R&D operations, hiring new staff across the board and employing contractors to upgrade systems both virtually and physically. The organisation is gaining a lot of attention for its recent developments and there is plenty of interest in the work currently being undertaken. At the same time, cyber threats are rapidly becoming realised, with highly-skilled attackers successfully penetrating well-respected organisations and operating for quite some time undetected.

*First Scenario:* An R&D employee working from home on their personal laptop notices their computer acting oddly. It is late at night, they are fatigued from a busy period at work, but they try everything they can think of to fix this issue themselves. Eventually they concede it is beyond their capabilities and retire for the night. The next morning when they clock on at work, they contact the organisation's help desk from their employee laptop, which works just fine. **Question:** How would you respond?

*Optional inject:* Upon escalation to the IR team, it is revealed the employee was the victim of a social engineering attack, where they clicked on an attachment in a spear phishing email to their work address. **Question:** Does this new information change your earlier response? If yes, how so?

*Optional inject:* The IR team now requires physical access to the compromised laptop. The employee is uncontactable and the laptop remains at their house. **Question:** How would you respond?

*Second Scenario:* It is the long weekend and few staff are at work when a number of servers stop responding. Calls are placed to the help desk from a selection of the skeleton staff, including a business executive, an R&D employee and a call centre worker, frustrated as they cannot do their job while the servers are offline. **Question:** How would you respond?

*Optional inject:* One of the server asset owners has not been contactable for three days. **Question:** Does this new information change your earlier response? Explain.

*Third Scenario:* The IR team has begun to notice an increase in volume and frequency of known, low-impact attacks. It is unclear what the reason for the increase is. **Question:** How would you respond?

*Optional inject:* The high frequency of attacks continues beyond 24 hours and level two analysts are assisting fatigued junior employees to manage the workload. Some staff members are beginning to become disgruntled at the extra "trivial" work and voice their grievances, not the least of which is missing the send-off afternoon tea for a valued staff member. **Question:** How would you respond?

*Fourth Scenario:* It is Friday afternoon – end of the work-week, when a large-scale distributed denial of service attack hits the organisation. IT systems are crippled, including communication channels such as email and mobile phones. **Question:** How would you respond? What takes first priority?

*Optional inject:* Systems are restored and services are beginning to be recovered across the organisation. **Question:** What happens next?